# A MODEL FOR THE DEGRADATION OF POLYIMIDES DUE TO OXIDATION

SATISH KARRA AND K. R. RAJAGOPAL

Abstract. Polyimides, due to their superior mechanical behavior at high temperatures, are used in a variety of applications that include aerospace, automobile and electronic packaging industries, as matrices for composites, as adhesives etc. In this paper, we extend our previous model in [S. Karra, K. R. Rajagopal, Modeling the non-linear viscoelastic response of high temperature polyimides, Mechanics of Materials, In press, doi:10.1016/j.mechmat.2010.09.006], to include oxidative degradation of these high temperature polyimides. Appropriate forms for the Helmholtz potential and the rate of dissipation are chosen to describe the degradation. The results for a specific boundary value problem, using our model compares well with the experimental creep data for PMR-15 resin that is aged in air.

## 1. Introduction

Polyimide and polyimide composites are used in a variety of applications due to the high glass transition temperature of above $300^oC$. These polymers and their composite components undergo degradation in a variety of ways including degradation due to oxidation. Thus, there is a need to understand how the mechanical behavior of these materials is affected by oxidation. Several experimental studies have been carried out which show that there is: (a) weight loss in the polyimides, and (b) an oxidized layer is formed on the surface of the material (see Bowles et al. (2001, 2003), also see references in Pochiraju and Tandon (2009)) due to oxidation. The loss of weight due to oxidation is observed to be due to chemical bond breakage and escape of volatile lower molecular weight gaseous products. In addition, it has been observed that the brittle oxidized layer formed on the surface of the polyimide acts as a crack initiation site, which leads to the failure of the materials. These cracks also provide more surface area for further degradation and damage due to oxidation. Recently, Tandon et al. (2006), Pochiraju and Tandon (2009), Roy et al. (2008) have looked at oxidative degradation of polymer composites from a modeling perspective. However, most of the works either do not consider the coupling between chemical reactions and deformation or assume that the coupling is between the small strain in a linearized elastic solid model (which does not correctly describe the mechanical behavior of these high temperature polymers since it has been experimentally shown that they exhibit non-linear viscoelastic response, see Falcone and Ruggles-Wrenn (2009)) and an advection-diffusion-reaction equation.

A thermodynamic framework that considers the coupling between chemical reactions (including stoichiometry and chemical kinetics) and deformation of polyimides that show non-linear viscoelastic response, is needed. Such a framework can also be used in modeling similar coupling in areas like asphalt mechanics, biomechanics and geo-mechanics. Some of the earlier works in areas of stoichiometry and thermochemistry are by Prigogine (1967), de Donder and Van Rysselberghe (1936), Van Rysselberghe (1958), Bowen (1968a,b), Samohỳl (1999), Nunziato and Walsh (1980), Björnbom (1975), Fishtik and Datta (2001), Germain et al. (1983), Pekar (2005), Zeleznik and Gordon (1968), and Kannan and Rajagopal (2010).

In this paper, we shall extend our constitutive theory that has been used to model the non-linear response of viscoelastic solids (see Karra and Rajagopal (2010)) to include degradation due to chemical reactions (specifically, oxidation). This theory is based on the thermodynamic framework of Rajagopal and co-workers (we refer the reader to Rajagopal and Srinivasa (2004) for details of this framework) that has been shown to be able to capture a plethora of phenomena. We extend our previous work by introducing a variable $\alpha$ that represents the extent of oxidation in the polyimide. Our approach should not be thought of as merely other internal variable theories that are in vogue; we are able to assign a clear meaning to this variable and thus it is a variable that goes towards specifying the state of the body. The forms for the Helmholtz $\psi$ potential and the rate of dissipation $\xi$ in Karra and Rajagopal (2010) are modified to incorporate the changes in the response of the body due to oxidative degradation. Our approach is similar to that of Rajagopal et al. (2007) who have modeled the degradation due to deformation and chain scission in polymers using a variable to quantify the degradation.

The current paper is organized as follows. The preliminaries that are required are documented in section (2). In sections (3.1), (3.2), the constitutive relations for the degradation due to oxidation are derived. In section (3.3), the predictions of our proposed model are compared with the experimental creep data for oxidative degradation of PMR-15 given in Ruggles-Wrenn and Broeckert (2009). In the final section, we make some remarks concerning the limitations of the approach, the scope for its improvement and future work that needs to be carried out.

## 2. Preliminaries

Let $\kappa_R(\mathcal{B})$ and $\kappa_t(\mathcal{B})$ denote the reference configuration and the current configuration, respectively. The motion $\boldsymbol{\chi}_{\kappa_R}$ is defined as the one-one mapping that assigns to each point $\boldsymbol{X} \in \kappa_R$, a point $\boldsymbol{x} \in \kappa_t$, at a time $t$, i.e.,

$$\boldsymbol{x} = \boldsymbol{\chi}_{\kappa_R}(\boldsymbol{X}, t). \tag{2.1}$$

The mapping $\chi_{\kappa_R}(\boldsymbol{X}, t)$ is assumed to be sufficiently smooth and invertible. Let $\kappa_{p(t)}$ be the configuration instantaneously reached by the body upon removal of the external stimuli. We shall call this configuration the natural configuration corresponding to $\kappa_t$ (see figure (1)). Let $\boldsymbol{F}$ be gradient of motion $\chi_{\kappa_R}(\boldsymbol{X}, t)$ (usually referred to as the deformation gradient), defined by

$$\boldsymbol{F} = \frac{\partial \boldsymbol{\chi}_{\kappa_R}}{\partial \boldsymbol{X}}, \tag{2.2}$$

and let the left and right Cauchy-Green tensors be defined through

$$\boldsymbol{B} = \boldsymbol{F}\boldsymbol{F}^T, \quad \boldsymbol{C} = \boldsymbol{F}^T\boldsymbol{F}. \tag{2.3}$$

The velocity is defined by

$$\boldsymbol{v} = \frac{\partial \boldsymbol{\chi}_{\kappa_R}}{\partial \boldsymbol{t}}. \tag{2.4}$$

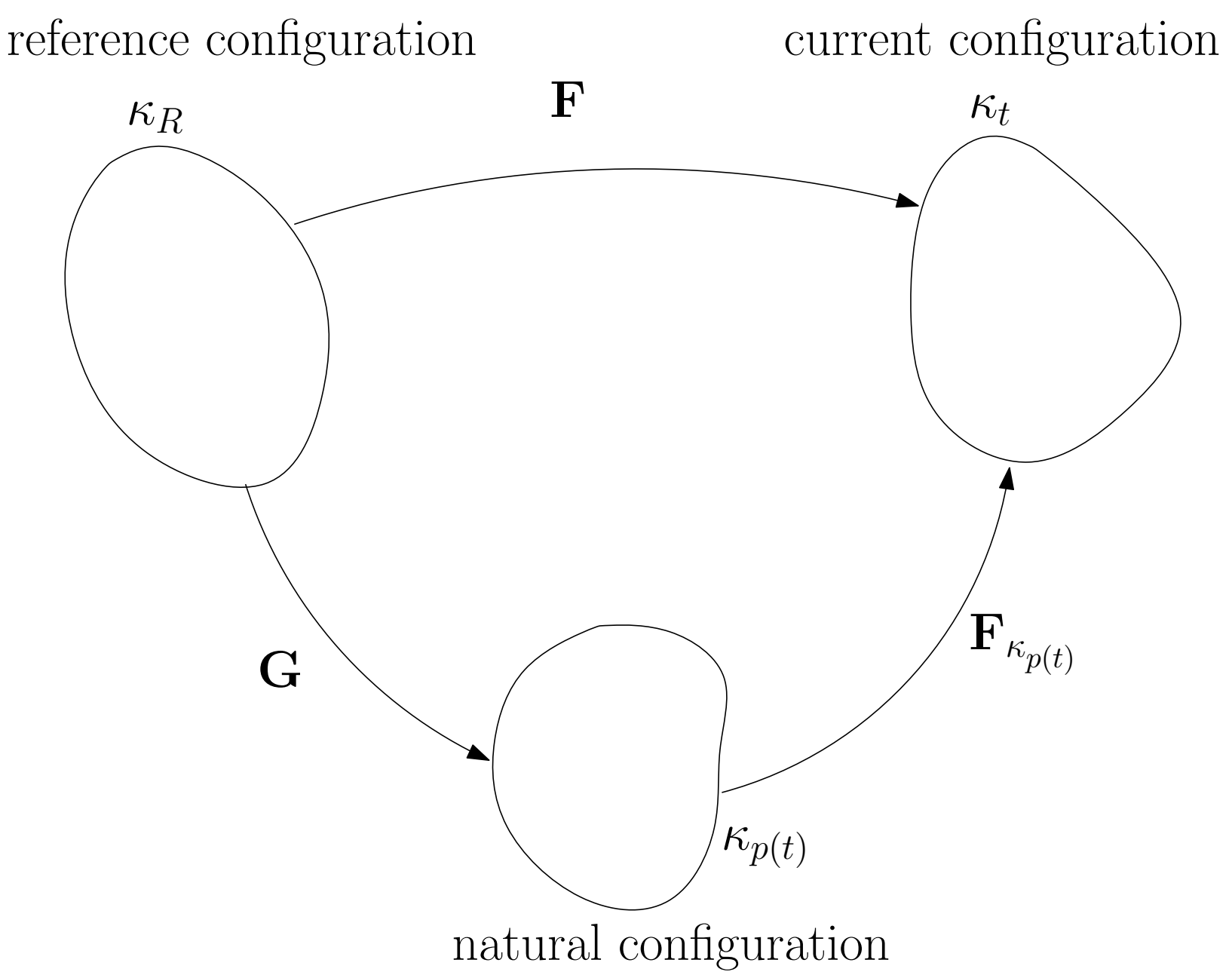

FIGURE 1. Illustration of the various configurations of the viscoelastic solid.

Let $\boldsymbol{F}_{\kappa_{p(t)}}$ be the gradient of the mapping from $\kappa_{p(t)}$ to $\kappa_t$[1] , and let $\boldsymbol{G}$ be defined by

$$\boldsymbol{G} = \boldsymbol{F}_{\kappa_R \to \kappa_{p(t)}} = \boldsymbol{F}_{\kappa_{p(t)}}^{-1} \boldsymbol{F}. \tag{2.5}$$

Similar to (2.3), we shall define the left Cauchy-Green stretch tensors $\boldsymbol{B}_G$ and $\boldsymbol{B}_{p(t)}$ through

$$\boldsymbol{B}_G = \boldsymbol{G}\boldsymbol{G}^T, \quad \boldsymbol{B}_{p(t)} = \boldsymbol{F}_{\kappa_{p(t)}} \boldsymbol{F}_{\kappa_{p(t)}}^T. \tag{2.6}$$

We shall also define the velocity gradients

$$\boldsymbol{L} = \dot{\boldsymbol{F}}\boldsymbol{F}^{-1}, \quad \boldsymbol{L}_G = \dot{\boldsymbol{G}}\boldsymbol{G}^{-1}, \quad \boldsymbol{L}_p = \dot{\boldsymbol{F}}_{\kappa_{p(t)}} \boldsymbol{F}_{\kappa_{p(t)}}^{-1}, \tag{2.7}$$

and their symmetric parts by

$$\boldsymbol{D}_i = \frac{1}{2}\left(\boldsymbol{L}_i + \boldsymbol{L}_i^T\right), \quad i = p(t), G \quad \text{or no subscript.} \tag{2.8}$$

Also, we define the principal invariants through

$$\mathrm{I}_{\boldsymbol{B}_l} = \mathrm{tr}(\boldsymbol{B}_l), \quad \mathrm{II}_{\boldsymbol{B}_l} = \frac{1}{2}\left\{[\mathrm{tr}(\boldsymbol{B}_l)]^2 - \mathrm{tr}(\boldsymbol{B}_l^2)\right\}, \quad \mathrm{III}_{\boldsymbol{B}_l} = \det(\boldsymbol{B}_l) \quad l = G, p(t), \tag{2.9}$$

where tr(.) is the trace operator for a second order tensor and det(.) is the determinant. Now, from (2.5)

$$\begin{aligned} \dot{\boldsymbol{F}} &= \dot{\boldsymbol{F}}_{\kappa_{p(t)}} \boldsymbol{G} + \boldsymbol{F}_{\kappa_{p(t)}} \dot{\boldsymbol{G}} \\ \Rightarrow \dot{\boldsymbol{F}}\boldsymbol{F}^{-1} &= \dot{\boldsymbol{F}}_{\kappa_{p(t)}} \boldsymbol{G}\boldsymbol{G}^{-1} \boldsymbol{F}_{\kappa_{p(t)}}^{-1} + \boldsymbol{F}_{\kappa_{p(t)}} \dot{\boldsymbol{G}} \\ \Rightarrow \boldsymbol{L} &= \boldsymbol{L}_{p(t)} + \boldsymbol{F}_{\kappa_{p(t)}} \boldsymbol{L}_G \boldsymbol{F}_{\kappa_{p(t)}}^{-1}, \end{aligned} \tag{2.10}$$

[1]In general, it is the mapping from the tangent space at a material point in $\kappa_{p(t)}$ to the tangent space at the same material point in $\kappa_t$.

where $\dot{(.)}$ is the material time derivative of the second order tensor. In addition,

$$\begin{aligned}\dot{\boldsymbol{B}}_{p(t)} &= \dot{\boldsymbol{F}}_{\kappa_{p(t)}}\boldsymbol{F}^T + \boldsymbol{F}\dot{\boldsymbol{F}}^T_{\kappa_{p(t)}} \\ &= \boldsymbol{L}_{p(t)}\boldsymbol{B}_{p(t)} + \boldsymbol{B}_{p(t)}\boldsymbol{L}^T_{p(t)},\end{aligned} \tag{2.11}$$

and similarly

$$\dot{\boldsymbol{B}}_G = \boldsymbol{L}_G\boldsymbol{B}_G + \boldsymbol{B}_G\boldsymbol{L}^T_G. \tag{2.12}$$

Hence, from (2.10) and (2.11), we have

$$\dot{\boldsymbol{B}}_{p(t)} = \boldsymbol{L}\boldsymbol{B}_{p(t)} + \boldsymbol{B}_{p(t)}\boldsymbol{L}^T_{p(t)} - \boldsymbol{F}_{\kappa_{p(t)}}\left(\boldsymbol{L}_G + \boldsymbol{L}^T_G\right)\boldsymbol{F}^T_{\kappa_{p(t)}}, \tag{2.13}$$

and so

$$\overset{\triangledown}{\boldsymbol{B}}_{p(t)} = -2\boldsymbol{F}_{\kappa_{p(t)}}\boldsymbol{D}_G\boldsymbol{F}^T_{\kappa_{p(t)}}, \tag{2.14}$$

where $\overset{\triangledown}{(.)}$ is the usual Oldroyd derivative defined through $\overset{\triangledown}{\boldsymbol{A}} := \dot{\boldsymbol{A}} - \boldsymbol{L}\boldsymbol{A} - \boldsymbol{A}\boldsymbol{L}^T$. The local form of the balance of mass, linear momentum, angular momentum (in the absence of internal couples), and energy are given by

$$\dot{\varrho} = -\varrho\,\mathrm{div}(\boldsymbol{v}), \tag{2.15a}$$

$$\varrho\dot{\boldsymbol{v}} = \mathrm{div}\left(\boldsymbol{T}^T\right) + \varrho\boldsymbol{b}, \tag{2.15b}$$

$$\boldsymbol{T} = \boldsymbol{T}^T, \tag{2.15c}$$

$$\varrho\dot{\epsilon} = \boldsymbol{T}.\mathbf{L} + \varrho r - \mathrm{div}(\boldsymbol{q}). \tag{2.15d}$$

where $\boldsymbol{T}$ is the Cauchy stress, $\varrho$ is the density, $\boldsymbol{b}$ is the specific body force, $\epsilon$ is the specific internal energy, $r$ is the radiant heating, $\boldsymbol{q}$ is the heat flux, div(.) stands for the divergence operator in the current configuration.

The local form of the second law of thermodynamics (for non-isothermal processes) is:

$$\boldsymbol{T}\cdot\boldsymbol{D} - \varrho\dot{\psi} - \varrho s\dot{\theta} - \frac{\boldsymbol{q}_h\cdot\mathrm{grad}(\theta)}{\theta} = \varrho\theta\zeta := \xi \geq 0, \tag{2.16}$$

where $\psi$ is the specific Helmholtz potential, $\varrho$ is the density, $\theta$ is the temperature, $s$ is the specific entropy, $\boldsymbol{q}_h$ is the heat flux, $\zeta$ is the rate of entropy production and $\xi$ is the rate of dissipation.

The kinematics presented in this section are sufficient for the work that follows.

## 3. Constitutive assumptions

3.1. **General results.** We shall assume that the viscoelastic solid is isotropic and incompressible with the specific Helmholtz potential of the form

$$\psi = \psi(\boldsymbol{B}_{p(t)}, \boldsymbol{B}_G, \theta, \alpha) = \hat{\psi}(\mathrm{I}_{\boldsymbol{B}_{p(t)}}, \mathrm{II}_{\boldsymbol{B}_{p(t)}}, \mathrm{I}_{\boldsymbol{B}_G}, \mathrm{II}_{\boldsymbol{B}_G}, \theta, \alpha), \tag{3.1}$$

where $\alpha$ is a variable that accounts for the extent of oxidation. $\alpha$ equal to zero implies the body is in its virgin state and $\alpha = 1$ means that the material is completely oxidized, and no further oxidation is possible. The rate of change of the variable $\alpha$ is related to the rate of the oxidation reaction that takes place in the polyimide. In general, $\boldsymbol{\alpha}$ can be taken to be a tensor which represents the degree of oxidation in different directions i.e., anisotropic oxidation. Although it is seen in experiments that oxidation mostly occurs on the surface of the polyimide, we shall assume that oxidation occurs at every point in the body. One can model the motion of the surface of oxidation in the polymide using a mixture theory approach such as that used by Rajagopal and Tao (1995) which can take into effect the diffusion of a singular surface, which in the case of the problem under consideration would be the surface that separates the region of the virgin and oxidized body; however such an approach would make the problem too complicated to be amenable to a meaningful study of a initial-boundary value problem. We are also assuming that oxygen, polyimide and the products of oxidation together constitute a constrained mixture i.e., there is no relative velocity between these constituents, and hence our approach does not capture the diffusion process. In order to model the diffusion phenomenon, one can follow the approach shown in Karra (2010).

Next, assuming that the elastic response from the current configuration $\kappa_t$ to the natural configuration $\kappa_{p(t)}$ is isotropic, without loss of generality, we choose $\kappa_{p(t)}$ such that

$$\boldsymbol{F}_{\kappa_{p(t)}} = \boldsymbol{V}_{\kappa_{p(t)}}, \tag{3.2}$$

where $\boldsymbol{V}_{\kappa_{p(t)}}$ is the right stretch tensor in the polar decomposition of $\boldsymbol{F}_{\kappa_{p(t)}}$. We shall also assume that the total rate of dissipation can be split additively as follows

$$\boldsymbol{T}\cdot\boldsymbol{D} - \varrho\dot{\psi} - \varrho s\dot{\theta} = \xi_{m,d} \geq 0, \quad -\frac{\boldsymbol{q}_h\cdot\mathrm{grad}(\theta)}{\theta} = \xi_c \geq 0, \tag{3.3}$$

where $\xi_{m,d}$ is the rate of dissipation due to the conversion of mechanical working into thermal energy and due to degradation, $\xi_c$ is the rate of dissipation due to heat conduction. Now, if we constitutively choose

$$\boldsymbol{q}_h = -\mathcal{K}(\theta)\,\mathrm{grad}(\theta), \quad \mathcal{K}(\theta) \geq 0, \tag{3.4}$$

where $\mathcal{K}$ is the thermal conductivity, then $(3.3)_{(b)}$ is automatically satisfied.

Now,

$$\begin{aligned}\dot{\psi} = &\left[\left(\frac{\partial\hat{\psi}}{\partial \mathrm{I}_{B_{p(t)}}} + \mathrm{I}_{B_{p(t)}}\frac{\partial\hat{\psi}}{\partial \mathrm{II}_{B_{p(t)}}}\right)\boldsymbol{I} - \frac{\partial\hat{\psi}}{\partial \mathrm{II}_{B_{p(t)}}}\boldsymbol{B}_{p(t)}\right]\cdot\dot{\boldsymbol{B}}_{p(t)} \\ &+ \left[\left(\frac{\partial\hat{\psi}}{\partial \mathrm{I}_{B_G}} + \mathrm{I}_{B_G}\frac{\partial\hat{\psi}}{\partial \mathrm{II}_{B_G}}\right)\boldsymbol{I} - \frac{\partial\hat{\psi}}{\partial \mathrm{II}_{B_G}}\boldsymbol{B}_G\right]\cdot\dot{\boldsymbol{B}}_G + \frac{\partial\hat{\psi}}{\partial\theta}\dot{\theta} + \frac{\partial\hat{\psi}}{\partial\alpha}\dot{\alpha},\end{aligned} \tag{3.5}$$

and using (2.11), (2.12) along with (3.2) in (3.5), we obtain that

$$\begin{aligned}\dot{\psi} = 2&\left[\left(\frac{\partial\hat{\psi}}{\partial \mathrm{I}_{B_{p(t)}}} + \mathrm{I}_{B_{p(t)}}\frac{\partial\hat{\psi}}{\partial \mathrm{II}_{B_{p(t)}}}\right)\boldsymbol{B}_{p(t)} - \frac{\partial\hat{\psi}}{\partial \mathrm{II}_{B_{p(t)}}}\boldsymbol{B}^2_{p(t)}\right]\cdot(\boldsymbol{D} - \boldsymbol{D}_G) \\ &+ 2\left[\left(\frac{\partial\hat{\psi}}{\partial \mathrm{I}_{B_G}} + \mathrm{I}_{B_G}\frac{\partial\hat{\psi}}{\partial \mathrm{II}_{B_G}}\right)\boldsymbol{B}_G - \frac{\partial\hat{\psi}}{\partial \mathrm{II}_{B_G}}\boldsymbol{B}^2_G\right]\cdot\boldsymbol{D}_G + \frac{\partial\hat{\psi}}{\partial\theta}\dot{\theta} + \frac{\partial\hat{\psi}}{\partial\alpha}\dot{\alpha}.\end{aligned} \tag{3.6}$$

Next, we shall assume the rate of dissipation $\xi_{m,d}$ to be of the form

$$\xi_{m,d} = \xi_{m,d}(\theta, \alpha, \dot{\alpha}, \boldsymbol{B}_{p(t)}, \boldsymbol{D}_G). \tag{3.7}$$

On substituting (3.6) into $(3.3)_{(a)}$, we arrive at

$$
\begin{aligned}
&\left[\boldsymbol{T} - 2\varrho\left(\frac{\partial\hat{\psi}}{\partial \mathrm{I}_{B_{p(t)}}} + \mathrm{I}_{B_{p(t)}}\frac{\partial\hat{\psi}}{\partial \mathrm{II}_{B_{p(t)}}}\right)\boldsymbol{B}_{p(t)} + 2\varrho\frac{\partial\hat{\psi}}{\partial \mathrm{II}_{B_{p(t)}}}\boldsymbol{B}^2_{p(t)}\right]\cdot\boldsymbol{D} \\
&+ 2\varrho\left[\left(\frac{\partial\hat{\psi}}{\partial \mathrm{I}_{B_{p(t)}}} + \mathrm{I}_{B_{p(t)}}\frac{\partial\hat{\psi}}{\partial \mathrm{II}_{B_{p(t)}}}\right)\boldsymbol{B}_{p(t)} - \frac{\partial\hat{\psi}}{\partial \mathrm{II}_{B_{p(t)}}}\boldsymbol{B}^2_{p(t)}\right]\cdot\boldsymbol{D}_G \\
&- 2\varrho\left[\left(\frac{\partial\hat{\psi}}{\partial \mathrm{I}_{B_G}} + \mathrm{I}_{B_G}\frac{\partial\hat{\psi}}{\partial \mathrm{II}_{B_G}}\right)\boldsymbol{B}_G - \frac{\partial\hat{\psi}}{\partial \mathrm{II}_{B_G}}\boldsymbol{B}^2_G\right]\cdot\boldsymbol{D}_G - \varrho\frac{\partial\hat{\psi}}{\partial\alpha}\dot{\alpha} \\
&- \varrho\left[\frac{\partial\hat{\psi}}{\partial\theta} + s\right]\dot{\theta} \\
&= \xi_{m,d}(\theta, \alpha, \dot{\alpha}, \boldsymbol{B}_{p(t)}, \boldsymbol{D}_G).
\end{aligned} \tag{3.8}
$$

We shall set

$$s = -\frac{\partial\hat{\psi}}{\partial\theta}, \tag{3.9}$$

and define

$$\boldsymbol{T}_{p(t)} := 2\varrho\left[\left(\frac{\partial\hat{\psi}}{\partial \mathrm{I}_{B_{p(t)}}} + \mathrm{I}_{B_{p(t)}}\frac{\partial\hat{\psi}}{\partial \mathrm{II}_{B_{p(t)}}}\right)\boldsymbol{B}_{p(t)} - \frac{\partial\hat{\psi}}{\partial \mathrm{II}_{B_{p(t)}}}\boldsymbol{B}^2_{p(t)}\right], \tag{3.10}$$

$$\boldsymbol{T}_G := 2\varrho\left[\left(\frac{\partial\hat{\psi}}{\partial \mathrm{I}_{B_G}} + \mathrm{I}_{B_G}\frac{\partial\hat{\psi}}{\partial \mathrm{II}_{B_G}}\right)\boldsymbol{B}_G - \frac{\partial\hat{\psi}}{\partial \mathrm{II}_{B_G}}\boldsymbol{B}^2_G\right]. \tag{3.11}$$

Using (3.9)–(3.11) in (3.8), we obtain

$$
\begin{aligned}
&\left(\boldsymbol{T} - \boldsymbol{T}_{p(t)}\right)\cdot\boldsymbol{D} + \left(\boldsymbol{T}_{p(t)} - \boldsymbol{T}_G\right)\cdot\boldsymbol{D}_G - \varrho\frac{\partial\hat{\psi}}{\partial\alpha}\dot{\alpha} \\
&\quad = \xi_{m,d}(\theta, \alpha, \dot{\alpha}, \boldsymbol{B}_{p(t)}, \boldsymbol{D}_G).
\end{aligned} \tag{3.12}
$$

By virtue of the constraint of incompressibility, we have

$$\mathrm{tr}(\boldsymbol{D}) = \mathrm{tr}(\boldsymbol{D}_{p(t)}) = \mathrm{tr}(\boldsymbol{D}_G) = 0. \tag{3.13}$$

Since, the right hand side of (3.12) does not depend on $\boldsymbol{D}$, using (3.13),

$$\boldsymbol{T} = p\boldsymbol{I} + \boldsymbol{T}_{p(t)}, \tag{3.14}$$

where $p$ is the Lagrange multiplier due to the constraint of incompressibility[2], with

$$\left(\boldsymbol{T}_{p(t)} - \boldsymbol{T}_G\right)\cdot\boldsymbol{D}_G - \varrho\frac{\partial\hat{\psi}}{\partial\alpha}\dot{\alpha} = \xi_{m,d}(\theta, \alpha, \dot{\alpha}, \boldsymbol{B}_{p(t)}, \boldsymbol{D}_G), \tag{3.15}$$

which can be re-written as

$$\left(\boldsymbol{T} - \boldsymbol{T}_G\right)\cdot\boldsymbol{D}_G - \varrho\frac{\partial\hat{\psi}}{\partial\alpha}\dot{\alpha} = \xi_{m,d}(\theta, \alpha, \dot{\alpha}, \boldsymbol{B}_{p(t)}, \boldsymbol{D}_G), \tag{3.16}$$

using (3.13) and (3.14).

[2]The standard method in continuum mechanics to obtain constraints appeals to the notion that the constraint response does not work. It has been shown recently by Rajagopal and Srinivasa (2005) that such an assumption is in general incorrect.

We shall further assume that $\xi_{m,d}$ can be further additively split as follows:

$$\xi_{m,d}(\theta, \dot{\alpha}, \boldsymbol{B}_{p(t)}, \boldsymbol{D}_G) = \xi_m(\theta, \alpha, \boldsymbol{B}_{p(t)}, \boldsymbol{D}_G) + \xi_d(\theta, \alpha, \dot{\alpha}), \tag{3.17}$$

with each of $\xi_m$, $\xi_d$ being non-negative, so that the second law is automatically satisfied. Noting that the first term and second terms on the left hand side of (3.16) are the contributions to dissipation[3] due to mechanical working and degradation, respectively, we shall further assume that

$$(\boldsymbol{T} - \boldsymbol{T}_G) \cdot \boldsymbol{D}_G = \xi_m(\theta, \alpha, \boldsymbol{B}_{p(t)}, \boldsymbol{D}_G), \tag{3.18a}$$

$$-\varrho \frac{\partial \hat{\psi}}{\partial \alpha} \dot{\alpha} = \xi_d(\theta, \alpha, \dot{\alpha}). \tag{3.18b}$$

Now, we shall maximize the rate of dissipation $\xi_m$ by varying $\boldsymbol{D}_G$ for fixed $\boldsymbol{B}_{p(t)}$. That is, we maximize the function

$$\Phi := \xi_m + \lambda_1 \left[\xi_m - (\boldsymbol{T} - \boldsymbol{T}_G) \cdot \boldsymbol{D}_G\right] + \lambda_2 (\boldsymbol{I} \cdot \boldsymbol{D}_G), \tag{3.19}$$

where $\lambda_1, \lambda_2$ are the Lagrange multipliers. By setting, $\partial\Phi/\partial\boldsymbol{D}_G = 0$, we get

$$\boldsymbol{T} = \boldsymbol{T}_G + \frac{\lambda_2}{\lambda_1}\boldsymbol{I} + \left(\frac{\lambda_1 + 1}{\lambda_1}\right) \frac{\partial \xi_m}{\partial \boldsymbol{D}_G}. \tag{3.20}$$

We need to determine the Lagrange multipliers. On substituting (3.20) into (3.16), we get

$$\left(\frac{\lambda_1 + 1}{\lambda_1}\right) = \frac{\xi_m}{\frac{\partial \xi_m}{\partial \boldsymbol{D}_G} \cdot \boldsymbol{D}_G}, \tag{3.21}$$

and so (3.20) with (3.11) becomes

$$\boldsymbol{T} = 2\varrho \left[ \left( \frac{\partial \hat{\psi}}{\partial \mathrm{I}_{B_G}} + \mathrm{I}_{B_G} \frac{\partial \hat{\psi}}{\partial \mathrm{II}_{B_G}} \right) \boldsymbol{B}_G - \frac{\partial \hat{\psi}}{\partial \mathrm{II}_{B_G}} \boldsymbol{B}_G^2 \right] + \left( \frac{\xi_m}{\frac{\partial \xi_m}{\partial \boldsymbol{D}_G} \cdot \boldsymbol{D}_G} \right) \frac{\partial \xi_m}{\partial \boldsymbol{D}_G} + \hat{\lambda} \boldsymbol{I}. \tag{3.22}$$

where $\hat{\lambda} := \frac{\lambda_2}{\lambda_1}$ is the Lagrange multiplier due to the constraint of incompressibility.

Finally, the constitutive relations are given by

$$\boldsymbol{T} = p\boldsymbol{I} + 2\varrho \left[ \left( \frac{\partial \hat{\psi}}{\partial \mathrm{I}_{B_{p(t)}}} + \mathrm{I}_{B_{p(t)}} \frac{\partial \hat{\psi}}{\partial \mathrm{II}_{B_{p(t)}}} \right) \boldsymbol{B}_{p(t)} - \frac{\partial \hat{\psi}}{\partial \mathrm{II}_{B_{p(t)}}} \boldsymbol{B}_{p(t)}^2 \right], \tag{3.23a}$$

$$\boldsymbol{T} = \hat{\lambda}\boldsymbol{I} + 2\varrho \left[ \left( \frac{\partial \hat{\psi}}{\partial \mathrm{I}_{B_G}} + \mathrm{I}_{B_G} \frac{\partial \hat{\psi}}{\partial \mathrm{II}_{B_G}} \right) \boldsymbol{B}_G - \frac{\partial \hat{\psi}}{\partial \mathrm{II}_{B_G}} \boldsymbol{B}_G^2 \right] + \left( \frac{\xi_m}{\frac{\partial \xi_m}{\partial \boldsymbol{D}_G} \cdot \boldsymbol{D}_G} \right) \frac{\partial \xi_m}{\partial \boldsymbol{D}_G}, \tag{3.23b}$$

$$\boldsymbol{q}_h = -k(\theta)\mathrm{grad}(\theta), \quad s = -\frac{\partial \hat{\psi}}{\partial \theta}, \tag{3.23c}$$

$$\varrho \frac{\partial \psi}{\partial \alpha} = -\frac{\xi_d}{\dot{\alpha}}. \tag{3.23d}$$

The two equations (3.23a), (3.23b) are to be equated and simplified to get the evolution equation for $\boldsymbol{B}_{\kappa_{p(t)}}$. This will be shown in the next subsection when we choose specific forms for $\psi$ and $\xi$.

[3]The term dissipation is used to refer to the mechanical working being converted into energy in thermal form, and associated with this dissipation we have entropy production. We shall abuse the use of the term dissipation and refer to other entropy producing mechanism such as degradation as also dissipation.

3.2. **Specific case.** We choose the specific Helmholtz potential as

$$\hat{\psi} = A^s + (B^s + c_2^s)(\theta - \theta_s) - \frac{c_1^s}{2}(\theta - \theta_s)^2 - c_2^s\theta \ln\left(\frac{\theta}{\theta_s}\right) + \frac{\mu_G(1 + \beta(\theta)\alpha)}{2\varrho}(\mathrm{I}_{B_G} - 3)$$
$$+ \frac{\mu_p(1 + \gamma(\theta)\alpha)}{2\varrho}(\mathrm{I}_{B_{p(t)}} - 3) + F(\alpha, \theta), \tag{3.24}$$

where $\mu_G, \mu_p$ are elastic parameters, $\theta_s$ is a reference temperature for the viscoelastic solid and the rates of mechanical dissipation, and the dissipation due to degradation as

$$\xi_m = \eta(1 + \delta(\theta)\alpha)\left(\boldsymbol{D}_G \cdot \boldsymbol{B}_{p(t)}\boldsymbol{D}_G\right), \tag{3.25a}$$
$$\xi_d = \frac{D\left(\|\dot{\alpha}\|\right)^{\frac{n+1}{n}}}{(1 - \alpha)^{\frac{1}{n}}}. \tag{3.25b}$$

where $\eta$ is the viscosity, $\|.\|$ stands for absolute value. Here, $\beta$, $\gamma$ and $\delta$ are material parameters that depend on temperature. Also, note from (3.25) that $\xi_m$, $\xi_d$ are non-negative provided $\eta$, $\delta$, $D$ are also non-negative.

Now,

$$s = -\frac{\partial \hat{\psi}}{\partial \theta}$$
$$= -(B^s + c_2^s) + c_1^s(\theta - \theta_s) + c_2^s \ln\left(\frac{\theta}{\theta_s}\right) + c_2^s - \frac{\mu_G \alpha}{2\varrho}\frac{\partial \beta}{\partial \theta}(\mathrm{I}_{B_G} - 3)$$
$$- \frac{\mu_p \alpha}{2\varrho}\frac{\partial \gamma}{\partial \theta}(\mathrm{I}_{B_{p(t)}} - 3) - \frac{\partial F}{\partial \theta}. \tag{3.26}$$

The internal energy $\epsilon$ is given by

$$\epsilon = \psi + \theta s$$
$$= A^s - B^s\theta_s + c_2^s(\theta - \theta_s) + \frac{c_1^s}{2}\left(\theta^2 - \theta_s^2\right) + \frac{\mu_G}{2\varrho}(\mathrm{I}_{B_G} - 3)\left[1 + \alpha\left(\beta - \theta\frac{\partial \beta}{\partial \theta}\right)\right] \tag{3.27}$$
$$+ \frac{\mu_p}{2\varrho}(\mathrm{I}_{B_{p(t)}} - 3)\left[1 + \alpha\left(\gamma - \theta\frac{\partial \gamma}{\partial \theta}\right)\right] + F - \theta\frac{\partial F}{\partial \theta}. \tag{3.28}$$

and the specific heat capacity $C_v$ is

$$C_v = \frac{\partial \epsilon}{\partial \theta} = c_1^s\theta + c_2^s - \frac{\mu_G \alpha \theta}{2\varrho}(\mathrm{I}_{B_G} - 3)\frac{\partial^2 \beta}{\partial \theta^2} - \frac{\mu_p \alpha \theta}{2\varrho}(\mathrm{I}_{B_{p(t)}} - 3)\frac{\partial^2 \gamma}{\partial \theta^2} - \theta\frac{\partial^2 F}{\partial \theta^2}. \tag{3.29}$$

Also, (3.23a), (3.23b) reduce to

$$\boldsymbol{T} = p\boldsymbol{I} + \bar{\mu}_p\boldsymbol{B}_{p(t)}, \tag{3.30a}$$
$$\boldsymbol{T} = \lambda\boldsymbol{I} + \bar{\mu}_G\boldsymbol{B}_G + \frac{\bar{\eta}}{2}\left(\boldsymbol{B}_{p(t)}\boldsymbol{D}_G + \boldsymbol{D}_G\boldsymbol{B}_{p(t)}\right), \tag{3.30b}$$

where $\bar{\mu}_p = \mu_p(1 + \beta(\theta)\alpha)$, $\bar{\mu}_G = \mu_G(1 + \gamma(\theta)\alpha)$, $\bar{\eta} = \eta(1 + \delta(\theta)\alpha)$. We also note that we chose the functions for the material moduli ($\bar{\mu}_p$, $\bar{\mu}_G$, $\bar{\eta}$) such that they increase as $\alpha$ goes from 0 to 1. This is consistent with the experiments (see figure 5 in Ruggles-Wrenn and Broeckert (2009)) where it seen that the *elastic modulus* increases with aging. We further note that such a choice of functions for the material moduli is different from what Rajagopal et al. (2007) have chosen in their work.

From (3.30)

$$(p-\lambda)\boldsymbol{I}+\bar{\mu}_p\boldsymbol{B}_{p(t)}=\bar{\mu}_G\boldsymbol{B}_G+\frac{\bar{\eta}}{2}\left(\boldsymbol{B}_{p(t)}\boldsymbol{D}_G+\boldsymbol{D}_G\boldsymbol{B}_{p(t)}\right), \tag{3.31}$$

and so by pre-multiplying the above equation by $\boldsymbol{B}_{p(t)}^{-1}$ and taking the trace, we get

$$(p-\lambda)=\frac{\bar{\mu}_G\mathrm{tr}(\boldsymbol{B}_{p(t)}^{-1}\boldsymbol{B}_G)-3\bar{\mu}_p}{\mathrm{tr}(\boldsymbol{B}_{p(t)}^{-1})}. \tag{3.32}$$

Using (3.32) in (3.31), we arrive at the following equation:

$$\left[\frac{\bar{\mu}_G\mathrm{tr}(\boldsymbol{B}_{p(t)}^{-1}\boldsymbol{B}_G)-3\bar{\mu}_p}{\mathrm{tr}(\boldsymbol{B}_{p(t)}^{-1})}\right]\boldsymbol{I}+\bar{\mu}_p\boldsymbol{B}_{p(t)}=\bar{\mu}_G\boldsymbol{B}_G+\frac{\bar{\eta}}{2}\left(\boldsymbol{B}_{p(t)}\boldsymbol{D}_G+\boldsymbol{D}_G\boldsymbol{B}_{p(t)}\right), \tag{3.33}$$

which can be re-written as

$$\begin{aligned}\left[\frac{\bar{\mu}_G\mathrm{tr}(\boldsymbol{B}_{p(t)}^{-1}\boldsymbol{B}_G)-3\bar{\mu}_p}{\mathrm{tr}(\boldsymbol{B}_{p(t)}^{-1})}\right]\boldsymbol{I}+\bar{\mu}_p\boldsymbol{B}_{p(t)}&=\bar{\mu}_G\boldsymbol{B}_G\\&-\frac{\bar{\eta}}{4}\left(\boldsymbol{V}_{p(t)}\overset{\triangledown}{\boldsymbol{B}}_{p(t)}\boldsymbol{V}_{\kappa_{p(t)}}^{-1}+\boldsymbol{V}_{\kappa_{p(t)}}^{-1}\overset{\triangledown}{\boldsymbol{B}}_{p(t)}\boldsymbol{V}_{p(t)}\right),\end{aligned} \tag{3.34}$$

where we have used (2.14) and (3.33). Now, (3.23d) reduces to

$$\frac{\mu_G\beta(\theta)}{2}(\mathrm{I}_{B_G}-3)+\frac{\mu_p\gamma(\theta)}{2}(\mathrm{I}_{B_{p(t)}}-3)+\varrho\frac{\partial F}{\partial\alpha}=-\frac{D\|\dot{\alpha}\|^{\frac{n+1}{n}}}{\dot{\alpha}\,(1-\alpha)^{\frac{1}{n}}}. \tag{3.35}$$

We shall assume that

$$F(\alpha)=-\frac{k(\theta)}{\varrho}\alpha, \tag{3.36}$$

where $k$ is a non-negative constant, then (3.35), for $n=1$ reduces to

$$\frac{\mu_G\beta(\theta)}{2}(\mathrm{I}_{B_G}-3)+\frac{\mu_p\gamma(\theta)}{2}(\mathrm{I}_{B_{p(t)}}-3)-k=-\frac{D\|\dot{\alpha}\|^2}{\dot{\alpha}\,(1-\alpha)}. \tag{3.37}$$

Notice that the first two terms on the left hand side of (3.37) represent the dependence of the extent of oxidation on the deformation of the material.

Thus, with the current choice of the specific Helmholtz potential and the rate of dissipation, we arrive at the following constitutive equations:

$$\boldsymbol{T}=p\boldsymbol{I}+\mu_p\left(1+\beta(\theta)\alpha\right)\boldsymbol{B}_{p(t)}, \tag{3.38}$$

where the evolution of the natural configuration is given by

$$\begin{aligned}&\left[\frac{\mu_G\left(1+\gamma(\theta)\alpha\right)\mathrm{tr}(\boldsymbol{B}_{p(t)}^{-1}\boldsymbol{B}_G)-3\mu_p\left(1+\beta(\theta)\alpha\right)}{\mathrm{tr}(\boldsymbol{B}_{p(t)}^{-1})}\right]\boldsymbol{I}+\mu_p\left(1+\beta(\theta)\alpha\right)\boldsymbol{B}_{p(t)}\\&=\mu_G\left(1+\gamma(\theta)\alpha\right)\boldsymbol{B}_G-\frac{\eta}{4}\left(\boldsymbol{V}_{p(t)}\overset{\triangledown}{\boldsymbol{B}}_{p(t)}\boldsymbol{V}_{\kappa_{p(t)}}^{-1}+\boldsymbol{V}_{\kappa_{p(t)}}^{-1}\overset{\triangledown}{\boldsymbol{B}}_{p(t)}\boldsymbol{V}_{p(t)}\right),\end{aligned} \tag{3.39}$$

and the evolution of $\alpha$ is given by (3.37). These constitutive relations reduce to the non-linear viscoelastic solid model derived in Karra and Rajagopal (2010) when there is no degradation. In a general initial-boundary value problem, one has to solve the coupled equations (2.15b), (3.37) along with (3.38), (3.39), subject to appropriate initial and boundary conditions. In problems where temperature gradients are important one needs to also consider the balance of energy (2.15d).

3.3. **Comparison with experimental data.** In order to compare the predictions of our model with experimental data, we shall consider the problem of uniaxial extension, given by

$$x = \lambda(t)X, \quad y = \frac{1}{\sqrt{\lambda(t)}}Y, \quad z = \frac{1}{\sqrt{\lambda(t)}}Z, \tag{3.40}$$

within the context of this model. The velocity gradient is given by

$$\boldsymbol{L} = \mathrm{diag}\left\{\frac{\dot{\lambda}}{\lambda}, -\frac{\dot{\lambda}}{2\lambda}, -\frac{\dot{\lambda}}{2\lambda}\right\}. \tag{3.41}$$

We shall assume that the stretch $\boldsymbol{B}_{p(t)}$ is given by

$$\boldsymbol{B}_{p(t)} = \mathrm{diag}\left\{B, \frac{1}{\sqrt{B}}, \frac{1}{\sqrt{B}}\right\}. \tag{3.42}$$

Straight forward calculations using (3.33) give

$$\frac{\dot{B}}{2} = \frac{B\dot{\lambda}}{\lambda} + \frac{\bar{\mu}_G}{\bar{\eta}}\frac{\lambda^2}{B} - \frac{\bar{\mu}_p}{\bar{\eta}}B - \left\{\frac{\frac{\bar{\mu}_G}{\bar{\eta}}\left(\lambda^3 + 2B^3\right) - 3\frac{\bar{\mu}_p}{\bar{\eta}}\lambda B^2}{\lambda B\left(1 + 2B^{3/2}\right)}\right\}, \tag{3.43}$$

which can be re-written in the following form:

$$\dot{\lambda} = \lambda\left\{\frac{\dot{B}}{2B} - \left[\frac{1}{\bar{\eta}B}\left(\bar{\mu}_G\frac{\lambda^2}{B} - \bar{\mu}_p B - \left(\frac{\bar{\mu}_G\left(\lambda^3 + 2B^3\right) - 3\bar{\mu}_p B^2\lambda}{B\lambda(1 + 2B^{3/2})}\right)\right)\right]\right\}. \tag{3.44}$$

Using (3.42) in (3.38), and also the fact that lateral surfaces are traction free, it is easy to see that

$$T_{11} = \bar{\mu}_p\left(B - \frac{1}{\sqrt{B}}\right), \tag{3.45}$$

and so

$$\dot{T}_{11} = \bar{\mu}_p\left(1 - \frac{1}{2\sqrt{B}}\right)\dot{B}. \tag{3.46}$$

In addition, (3.37) becomes

$$-\frac{\mu_G\beta(\theta)}{2}\left(\frac{\lambda^2}{B} - \frac{2\sqrt{B}}{\lambda} - 3\right) - \frac{\mu_p\gamma(\theta)}{2}\left(B + \frac{2}{\sqrt{B}} - 3\right) + k = \frac{D\|\dot{\alpha}\|^2}{\dot{\alpha}\left(1 - \alpha\right)}. \tag{3.47}$$

We shall also use logarithmic strain (or true strain) $\varepsilon = \ln\lambda$ as our strain measure in what follows.

We shall compare our model with the experimental data for PMR-15 from Ruggles-Wrenn and Broeckert (2009). With the given $\dot{T}_{11}$ and material parameters, (3.46), (3.44) were first solved using the initial condition that $B(0) = \lambda(0) = 1$ for a time of $\frac{T_{11}}{\dot{T}_{11}}$. Then, (3.44), (3.45) were solved till the end of loading. Since in the experiments the aging in air was done without any load being applied, (3.37) was also solved without the first two terms on the left hand side using $\alpha(0) = 0$ as the initial condition. The ODEs were solved in MATLAB using the `ode45` solver. The following parameters were used for comparing the results predicted by our model to the experimental creep data for PMR-15 (under 10 MPa loading) that has been aged in air for various amounts of time (also see figure (2)): $\mu_p = 2 \times 10^9$ MPa, $\mu_G = 3.8 \times 10^8$ MPa, $\eta = 45 \times 10^{12}$ MPa.s, $\frac{k}{D} = 1.2 \times 10^{-6}$ s$^{-1}$, $\beta = 10$, $\gamma = 0.3$, $\delta = 0.5$. Also, since the experiments were done under isothermal conditions, the balance of energy (2.15d) was not considered.

## 4. Concluding remarks

A model for the degradation of polyimides due to oxidation has been developed in this paper. Our model also accounts for the effect of deformation on the aging due to oxidation. However, there is no experimental data to corroborate this part of our model.

The limitations of our model are as follows:

(1) Our model cannot predict the diffusion of oxygen and hence one cannot estimate the thickness of the oxygen layer on the surface of the resin.
(2) The weight lost due to oxidation cannot be estimated using our model. For this one needs to understand the chemical kinetics. Once the chemical kinetics are established, an approach similar to ours can be used to couple these reaction kinetics to the deformation of the polymer. Our work in this paper can be viewed as a first step towards this end.

Directions for future work are as follows:

(1) Further experiments need to be carried out to get the creep and stress relaxation of different polyimides under different loading conditions and for various amounts of aging in oxygen. Also, one needs to perform experiments where simultaneous deformation and aging takes place to see how they are coupled.
(2) One can easily extend our work to include anisotropy in oxidation by introducing a tensor ($\boldsymbol{\alpha}$), and also anisotropic response of the body by modifying the Helmholtz potential. In addition, for modeling oxidative degradation in polyimide composites, one can also use different variables $\alpha_m$, $\alpha_f$ that represent the degradation due to oxidation in the polymer matrix and fiber, respectively. This approach is similar to Baek and Pence (2009) who have used different variables to represent the degradation due to swelling in the composite matrix and fiber.
(3) Our framework in this paper can also be used to model oxidation or other chemical reactions in materials like asphalt/asphalt derivatives, and biomaterials. For this, one has to accordingly modify the terms in the Helmholtz potential and the rate of dissipation due to deformation.

## Acknowledgements

The authors thank AFOSR/AFRL for supporting this work. Part of this work was done when Satish Karra was appointed as a lecturer during his Ph.D. by the Department of Mechanical Engineering at Texas A&M University. He appreciates this support by the department.

## References

Baek, S. and T. J. Pence (2009). On swelling induced degradation of fiber reinforced polymers. *International Journal of Engineering Science 47*(11-12), 1100–1109.

Björnbom, P. H. (1975). The independent reactions in calculations of complex chemical equilibria. *Industrial & Engineering Chemistry Fundamentals 14*(2), 102–106.

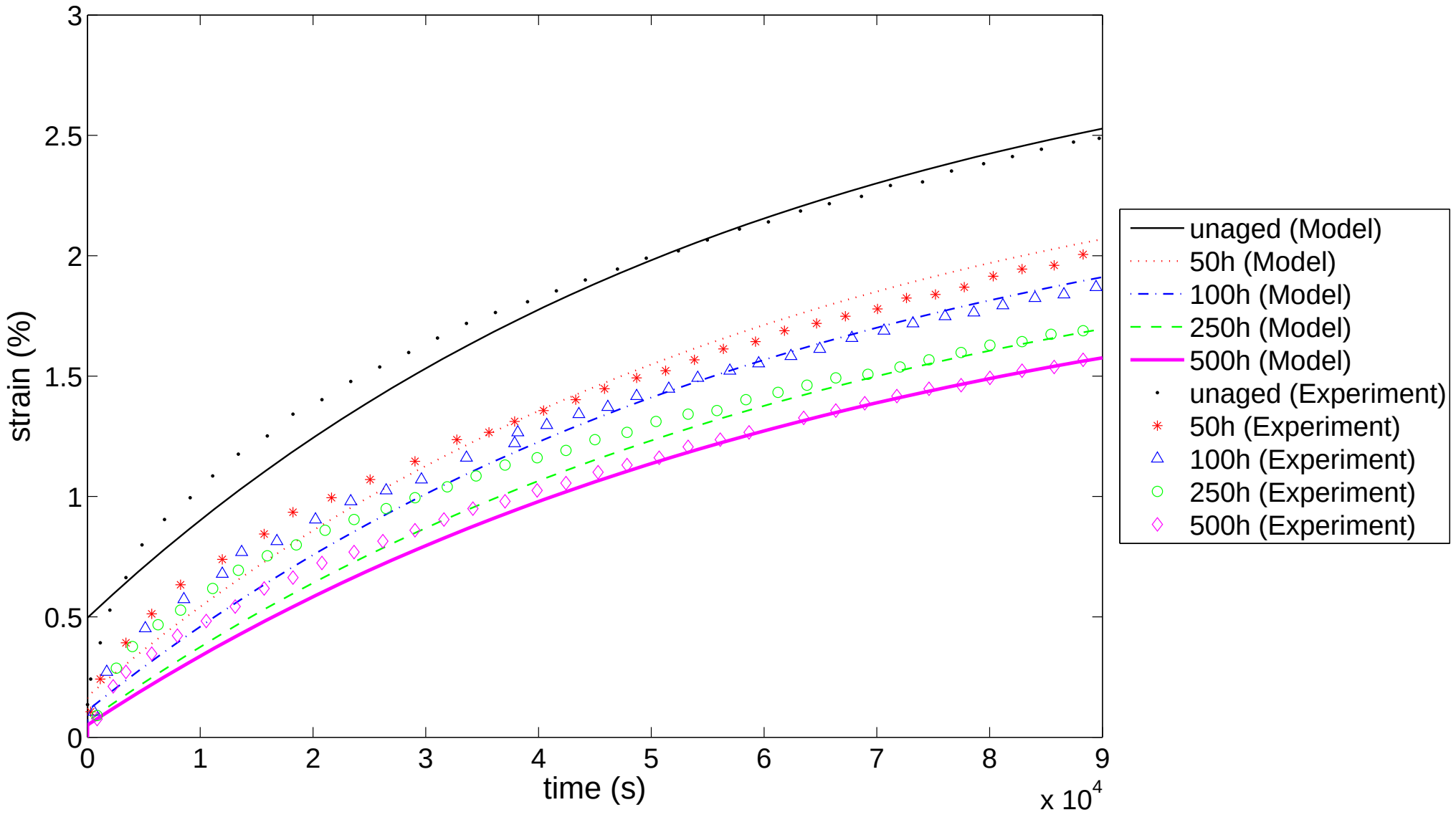

FIGURE 2. Comparison of the model predictions with experimental creep data by Ruggles-Wrenn and Broeckert (2009) for PMR-15 at a loading of 10 MPa with $\dot{T}_{11} = 1$ MPa/s as the rate of loading. The amount of time that the sample is aged in air is also shown. The material parameters used are as follows: $\mu_p = 2 \times 10^9$ MPa, $\mu_G = 3.8 \times 10^8$ MPa, $\eta = 45 \times 10^{12}$ MPa.s, $\frac{k}{D} = 1.2 \times 10^{-6}$ s$^{-1}$, $\beta = 10$, $\gamma = 0.3$, $\delta = 0.5$.

Bowen, R. M. (1968a). On the stoichiometry of chemically reacting materials. *Archive for Rational Mechanics and Analysis 29*(2), 114–124.

Bowen, R. M. (1968b). Thermochemistry of reacting materials. *The Journal of Chemical Physics 49*, 1625–1637.

Bowles, K. J., D. S. Papadopoulos, L. L. Inghram, L. S. McCorkle, and O. V. Klan (2001). Longtime durability of pmr-15 matrix polymer at 204, 260, 288 and $316^o c$. Technical Report 210602, NASA.

Bowles, K. J., L. Tsuji, J. Kamvouris, and G. D. Roberts (2003). Long-Term Isothermal Aging Effects on Weight Loss, Compression Properties, and Dimensions of T650-35 Fabric-Reinforced PMR-15 Composites–Data. Technical Report 211870, NASA.

de Donder, T. and P. Van Rysselberghe (1936). *Thermodynamic Theory of Affinity*. Palo Alto: Stanford University Press.

Falcone, C. M. and M. B. Ruggles-Wrenn (2009). Rate dependence and short-term creep behavior of a thermoset polymer at elevated temperature. *Journal of Pressure Vessel Technology 131*, 011403(1–8).

Fishtik, I. and R. Datta (2001). De Donder relations in mechanistic and kinetic analysis of heterogeneous catalytic reactions. *Ind. Eng. Chem. Res 40*(11), 2416–2427.

Germain, P., P. Suquet, and Q. S. Nguyen (1983). Continuum thermodynamics. *ASME, Transactions, Journal of Applied Mechanics 50*(4), 1010–1020.

Kannan, K. and K. R. Rajagopal (2010). A thermodynamical framework for chemically reacting systems. *Zeitschrift für Angewandte Mathematik und Physik (ZAMP) In Press, Accepted Manuscript*.

Karra, S. (2010). Diffusion of a fluid through a viscoelastic solid. *Arxiv preprint arXiv:1010.3488*.

Karra, S. and K. R. Rajagopal (2010). Modeling the non-linear viscoelastic response of high temperature polyimides. *Mechanics of Materials In Press, Accepted Manuscript*.

Nunziato, J. W. and E. K. Walsh (1980). On ideal multiphase mixtures with chemical reactions and diffusion. *Archive for Rational Mechanics and Analysis 73*(4), 285–311.

Pekar, M. (2005). Thermodynamics and foundations of mass-action kinetics. *Progress in Reaction Kinetics and Mechanism, 30 1*(2), 3–113.

Pochiraju, K. and G. P. Tandon (2009). Interaction of oxidation and damage in high temperature polymeric matrix composites. *Composites Part A: Applied Science and Manufacturing 40*(12), 1931–1940.

Prigogine, I. (1967). *Introduction to Thermodynamics of Irreversible Processes*. New York: John Wiley & Sons.

Rajagopal, K. R. and A. R. Srinivasa (2004). On the thermomechanics of materials that have multiple natural configurations Part I: Viscoelasticity and classical plasticity. *Zeitschrift für Angewandte Mathematik und Physik (ZAMP) 55*(5), 861–893.

Rajagopal, K. R. and A. R. Srinivasa (2005). On the nature of constraints for continua undergoing dissipative processes. *Proceedings of the Royal Society A: Mathematical, Physical and Engineering Science 461*(2061), 2785–2795.

Rajagopal, K. R., A. R. Srinivasa, and A. S. Wineman (2007). On the shear and bending of a degrading polymer beam. *International Journal of Plasticity 23*(9), 1618–1636.

Rajagopal, K. R. and L. Tao (1995). *Mechanics of Mixtures*. Singapore: World Scientific.

Roy, S., S. Singh, and G. A. Schoeppner (2008). Modeling of evolving damage in high temperature polymer matrix composites subjected to thermal oxidation. *Journal of Materials Science 43*(20), 6651–6660.

Ruggles-Wrenn, M. B. and J. L. Broeckert (2009). Effects of prior aging at 288$^o$C in air and in argon environments on creep response of PMR-15 neat resin. *Journal of Applied Polymer Science 111*(1), 228–236.

Samohỳl, I. (1999). Thermodynamics of reacting mixtures of any symmetry with heat conduction, diffusion and viscosity. *Archive for Rational Mechanics and Analysis 147*(1), 1–45.

Tandon, G. P., K. V. Pochiraju, and G. A. Schoeppner (2006). Modeling of oxidative development in PMR-15 resin. *Polymer Degradation and Stability 91*(8), 1861–1869.

Van Rysselberghe, P. (1958). Reaction rates and affinities. *The Journal of Chemical Physics 29*, 640–642.

Zeleznik, F. J. and S. Gordon (1968). Calculation of complex chemical equilibria. *Industrial & Engineering Chemistry 60*(6), 27–57.

Satish Karra (Corresponding Author), Texas A&M University, Department of Mechanical Engineering, 3123 TAMU, College Station TX 77843-3123, United States of America
*E-mail address*: satkarra@tamu.edu

K. R. Rajagopal, Texas A&M University, Department of Mechanical Engineering, 3123 TAMU, College Station TX 77843-3123, United States of America
*E-mail address*: krajagopal@tamu.edu